\documentclass[a4paper, oneside, twocolumn, notitlepage, 10pt]{extarticle_ecoc}
\usepackage{ecoc}
\newcommand{\abs}[1]{\lvert#1\rvert}

\begin{document}
\selectlanguage{english}    


\title{Complex-Valued Kernel-based Phase and Amplitude Distortion Compensation in Parametrically Amplified Optical Links}%


\author{
    Long H. Nguyen\textsuperscript{(1)}, Sonia Boscolo\textsuperscript{(1)},
    Stylianos Sygletos\textsuperscript{(1)}
}

\maketitle                  


\begin{strip}
    \begin{author_descr}

        \textsuperscript{(1)} Aston Institute of Photonic Technologies, Aston University, Birmingham B4 7ET, United Kingdom
        \textcolor{blue}{\uline{lnguy19@aston.ac.uk}}



    \end{author_descr}
\end{strip}

\renewcommand\footnotemark{}
\renewcommand\footnoterule{}


\begin{strip}
    \begin{ecoc_abstract}
 We develop a complex-valued kernel-adaptive-filtering based method for phase and amplitude distortion compensation in cascaded fibre-optical parametric amplifier (FOPA) links. Our algorithm predicts and cancels both distortions induced by pump-phase modulation across all amplification stages, achieving more than an order of magnitude improvement in BER. \textcopyright2024 The Author(s)
    \end{ecoc_abstract}
\end{strip}


\section{Introduction}
Fibre-optical parametric amplifiers (FOPAs) offer unique properties that make them attractive for future optical communications, including their ability to operate at nearly any frequency with significant gain bandwidth, ultra-fast response, and capabilities for wavelength conversion and optical phase conjugation (OPC)\cite{marhic_book2007}. However,  stimulated Brillouin scattering (SBS) poses a major challenge to integrating FOPAs into optical systems\cite{gordyenko_oft2021} as it limits the pump power deliverable to the highly nonlinear fibre (HNLF), thereby restricting the achievable signal gain. To mitigate SBS, a common strategy is to broaden the pump source's linewidth via external phase modulation, e.g. using a set of radio-frequency (RF) tones\cite{coles_oe2010}, which effectively suppresses SBS and enables higher gains \cite{torounidis_ptl2006}. However, this modulation can introduce temporal fluctuations in the complex parametric gain, impacting  coherently-detected complex amplitude signals

Recent research has addressed SBS mitigation at the subsystem and device level, showing promising results but often resulting in more complex device architectures which could hinder commercialization due to increased costs\cite{bastamova_oft2024}. Conversely, integrating machine learning (ML) into DSP has shown potential for enhancing equalization functionality in commercial systems. Specifically, ML-enhanced DSP tailored for FOPAs could offer a viable solution for their broader adoption in communication systems. Our recent efforts have developed DSP techniques that effectively mitigate phase distortion from pump dithering in FOPA devices \cite{boscolo_oe2022, nguyen_oe2024}.  
 
Kernel-based online learning algorithms, often referred to as kernel adaptive filters, leverage the theory of reproducing kernel Hilbert spaces (RKHSs) for nonlinear prediction in real-time applications. These methods transform inputs to a high-dimensional feature space through a positive definite kernel function, allowing nonlinear optimization problems in the input space to be treated as convex optimization problems in RKHS \cite{rojo-alvarez2018}.  In\cite{boscolo_oe2022}, we have successfully applied the sliding-window (SW) version of the real-valued kernel recursive least-squares (KRLS) adaptive algorithm introduced in\cite{vaerenbergh2006} for mitigating the dithering-induced phase distortion in OPC systems following conventional carrier phase recovery (CPR)\cite{fatadin_jlt2009}. 

In this paper, we extend the operation of the SWKRLS algorithm to the complex domain to address also the related amplitude distortions in cascaded FOPA links. Our complex-valued (CV) algorithm functions as a single-stage online equaliser handling both amplitude and phase distortions from pump dithering and random phase noise induced from laser sources, without needing prior knowledge of the dithering frequencies unlike the method demonstrated in\cite{nguyen_oe2024}. The performance of the equaliser is verified numerically in 28-Gbaud single-polarisation 16 quadrature-amplitude modulation (QAM) signal transmission over a cascade of ten FOPAs, achieving substantial bit-error-rate (BER) improvement over conventional CPR. \vspace{-0.3cm}

\section{Method}
\begin{figure*}[t!]
    \centering
    \includegraphics[width=15cm]{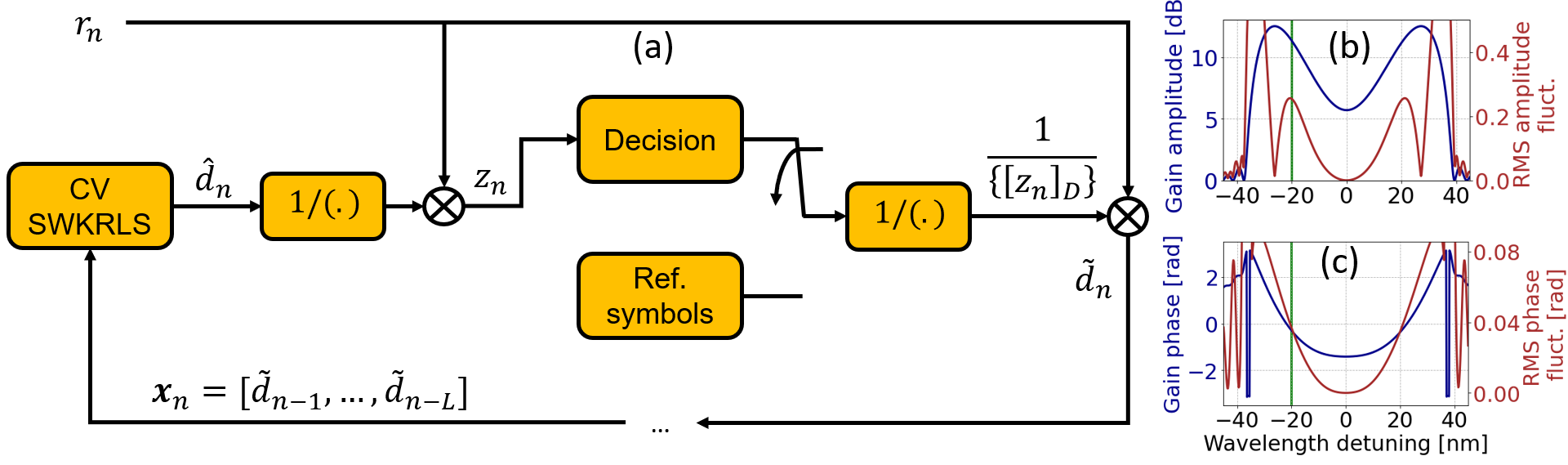}
    \caption{(a) Block diagram of the proposed CV SWKRLS-based equaliser for $n$th symbol recovery. (b-c) Amplitude and phase spectral responses of the FOPA's gain versus wavelength detuning from the pump, and respective dithering-induced RMS fluctuations. The green vertical line indicates the operating point for the performance evaluation of the proposed algorithm.}
    \label{fig:figure1}
\end{figure*}

Our proposed scheme is illustrated in Fig. \ref{fig:figure1}(a) while Fig. \ref{fig:figure1}(b-c) displays the complex gain response of each FOPA unit and the corresponding root-mean-square (RMS) fluctuations induced by dithering as a function of the signal detuning from the pump wavelength. At the point of maximum gain, the signal primarily experiences phase distortion, while smaller wavelength detunings introduce also amplitude distortion. 


The accumulated impact of the aforementioned distortions appears as an unknown complex factor on the received signal $r_n$, where $n$ represents the symbol index. At the core of our compensation scheme is the complex valued SWKRLS (CV-SWKRLS) algorithm that creates an estimate of this distortion, denoted as $\hat{d}_n$, subsequently used for the compensation purpose. The corrected symbol $z_n = r_n / \hat{d}_n$ is then processed through a decision circuit, which outputs the decision-directed symbol $[z_n]_D$. The complex ratio of the received symbol to the decision-directed symbol $\tilde{d}_n = r_n / [z_n]_D$, represents a decision-driven distortion, which has critical role in updating our algorithm and maintaining its online operation. A set of reference symbols is utilised for the initial training of the model until its convergence before the algorithm enters a decision-directed mode.

A way to operate kernel algorithms in the complex domain is through "complexification" of real RKHS \cite{Bouboulis_TSP2011}. In this case, the input complex data $\mathbf{x} = \mathbf{x_r} + j \mathbf{x_i}$, where $\mathbf{x_r},\mathbf{x_i} \in \mathbb{R}^L$ is mapped to a "complexified" RKHS according to the rule $\mathbf{\hat{\Phi}}(\mathbf{x})=\Phi(\mathbf{x_r}, \mathbf{x_i})+j\Phi(\mathbf{x_r}, \mathbf{x_i})$, where $\Phi(\mathbf{x_r}, \mathbf{x_i}) = \kappa(\cdot, (\mathbf{x_r}, \mathbf{x_i}))$ is the feature map of a chosen real kernel, which in our case is Gaussian, i.e.  $\kappa(\mathbf{p}, \mathbf{q}) := \mathrm{exp}[-\sum_{k=1}^{2L}(p_k-q_k)^2/(2\sigma^2)]$, where $\mathbf{p},\mathbf{q} \in \mathbb{R}^{2L}$ and the hyperparameter $\sigma$ defines the kernel width. In the regression task between the input $\hat{\mathbf{X}} \in \mathbb{R}^{M \times 2L}$ (consisting of real and imaginary parts of the complex data points) and the output $\mathbf{y} \in \mathbb{C}^{M}$ where $M$ is the number of observations, the algorithm aims to achieve $\min_{\mathbf{\alpha} \in \mathbb{C}^M}||\mathbf{y}-\mathbf{K}\mathbf{\alpha}||^2$, where $\mathbf{K}$ is the kernel matrix calculated on the input observations, i.e., $\mathbf{K}(p, q)=\kappa(\hat{\mathbf{X}}_p, \hat{\mathbf{X}}_q)$, with $\hat{\mathbf{X}}_p$ and $\hat{\mathbf{X}}_q$ are $p$th and $q$th rows of $\hat{\mathbf{X}}$ respectively. The coefficient vector $\mathbf{\alpha}$ can be solved directly but in the online scenario, a recursive update through a SW approach will reduce the complexity.

The online SWKRLS algorithm will make prediction of the one-step ahead complex distortion $\hat{d}_{n}$ based on a vector of past $L$ decision-driven distortions $\mathbf{x}_{n}=[\tilde{d}_{n-1}, ..., \tilde{d}_{n-L}]$, which at the input of the algorithm is converted at to a real valued composite representation $\hat{\mathbf{x}}_{n} = [\mathfrak{Re}(\mathbf{x}_{n}), \mathfrak{Im}(\mathbf{x}_{n})] \in \mathbb{R}^{2L}$, to facilitate calculation with real Gaussian kernels, and the learning coefficient vector $\mathbf{\alpha}_{n-1}$ carried from the previous update. 
After obtaining the new decision-driven distortion $\tilde{d}_n$ from the recently predicted $\hat{d}_n$, in the CV SWKRLS algorithm, we only consider the last $M$ input-output pairs as observations and form the observation input matrix $\hat{\mathbf{X}}_n = [\hat{\mathbf{x}}_{n}, ..., \hat{\mathbf{x}}_{n-M+1}]^\mathrm{T}$ and the observation output vector $\mathbf{y}_n=[\tilde{d}_{n}, ..., \tilde{d}_{n-M+1}]^\mathrm{T}$. The regularised real kernel matrix can be defined as $\mathbf{K}_n = \hat{\mathbf{X}}_n\hat{\mathbf{X}}_n^{\mathrm{T}}+\lambda\mathbf{I}$ ($\mathbf{I}$ is the identity matrix and $\lambda$ is a regularisation constant). The updated solution is then obtained as $\mathbf{\alpha}_n=\mathbf{K}_n^{-1}\mathbf{y}_n$. However, this equation requires recalculation of the inverse kernel matrix at every symbol update and it is very computationally expensive and memory-intensive. Therefore, instead of explicitly computing $\mathbf{K}_n$ and then $\mathbf{K}_n^{-1}$ every time, these matrices can be updated recursively by leveraging the matrices from the previous cycle, $\mathbf{K}_{n-1}$ and $\mathbf{K}_{n-1}^{-1}$\cite{vaerenbergh2006}. Specifically, the new regularised kernel matrix $\mathbf{K}_n$ is constructed from $\mathbf{K}_{n-1}$, by removing its first row and column, now referred to as $\mathbf{\bar{K}}_{n-1}$, and then adding kernels calculated with the new data as the last row and column:
\begin{equation}
\label{eq:K_n}
    \mathbf{K}_n=
    \begin{bmatrix}
    \mathbf{\bar{K}}_{n-1} & \mathbf{b}_n \\
    \mathbf{b}_n^{\mathrm{T}} & c_n
    \end{bmatrix},
\end{equation}
where $\mathbf{b}_n=[\kappa(\hat{\mathbf{x}}_{n-M+1}, \hat{\mathbf{x}}_n), ..., \kappa(\hat{\mathbf{x}}_{n-1}, \hat{\mathbf{x}}_n)]^{\mathrm{T}}$, and $c_n=\kappa(\hat{\mathbf{x}}_n, \hat{\mathbf{x}}_n)+\lambda$.  The inverse kernel matrix $\mathbf{K}_n^{-1}$ can then be updated accordingly as
\begin{equation}
\label{eq:inverse K_n}
    \mathbf{K}_n^{-1}=
    \begin{bmatrix}
    \mathbf{\bar{K}}_{n-1}^{-1}-\mathbf{\bar{K}}_{n-1}^{-1}\mathbf{b}_n\mathbf{e}_n^{\mathrm{T}} & \mathbf{e}_n \\
    \mathbf{e}_n^{\mathrm{T}} & f_n
    \end{bmatrix},
\end{equation}

where $f_n=(c_n-\mathbf{b}_n^{\mathrm{T}}\mathbf{\bar{K}}_{n-1}^{-1}\mathbf{b}_n)^{-1}$ and $\mathbf{e}_n=-\mathbf{\bar{K}}_{n-1}^{-1}\mathbf{b}_n f_n$. After calculating the updated solution $\mathbf{\alpha}_n$, our estimation algorithm, in the next cycle, can predict distortion for the $(n+1)$th symbol as $\hat{d}_{n+1} = \mathbf{h}_{n+1}^\mathrm{T} \alpha_n$ by applying the learned $\mathbf{\alpha}_n$ vector on the kernel vector evaluated on the points within the dictionary $\hat{\mathbf{X}}_n$ and the new point $\hat{\mathbf{x}}_{n+1}$, i.e., $\mathbf{h}_{n+1} =[\kappa(\hat{\mathbf{x}}_{n-M+1}, \hat{\mathbf{x}}_{n+1}), \dots, \kappa(\hat{\mathbf{x}}_n, \hat{\mathbf{x}}_{n+1})]^\mathrm{T}$.


\section{Results and discussion}

\begin{figure}[t!]
    \centering
    \includegraphics[width=7cm]{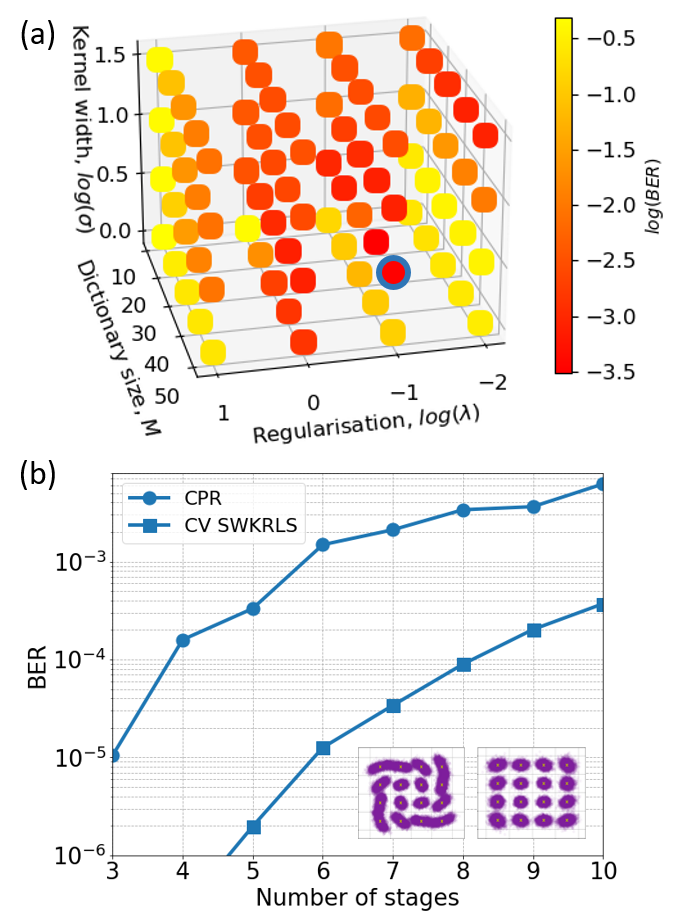}
    \caption{(a) Hyyper-parameter optimisation with input block length $L = 20$. The blue circled point represents the selection for the subsequent study. (b) BER performance comparison over number of cascaded FOPA stages. }
    \label{fig:figure2}
\end{figure}

We considered a system comprising ten identical cascaded stages, each incorporating linear loss followed by a FOPA. We modelled the FOPA using the complex signal gain\cite{marhic_book2007} with the phase mismatch comprised an additional instantaneous term induced by the phase modulation of the pump\cite{mussot_ptl2004}. Our pump-phase  modulation employed a four-tone scheme with a base frequency of $100\,$MHz and a multiple of three spacing between successive tones, i.e., the set of RF tones was $[0.1,\, 0.3,\, 0.9,\, 2.7]\,$GHz. The tone’s amplitudes and phases were optimised to ensure nearly uniform power distribution among the peaks across the broadened pump spectrum. This optimisation was performed using stochastic gradient descent in TensorFlow\cite{nguyen_oe2024}. The optimised pump-phase modulation significantly raised the SBS power threshold, allowing for SBS-free operation of the FOPA at its peak power gain of $25\,$dB\cite{nguyen_oe2024}. To illustrate the equalisation capability of our kernel method, the FOPA was operated at a signal detuning from the pump of $\abs{\lambda_s-\lambda_p}=20\,$nm (see Fig.\ \ref{fig:figure1}(b-c)). 

We performed numerical simulations of the transmission of a single-polarisation 28-Gbaud 16-QAM Nyquist shaped signal with a roll-off factor of $0.1$. The laser line-widths were $50\,$kHz and $30\,$kHz for the transmitter and receiver units and the FOPA pumps, respectively. To avoid the signal symbols experiencing exactly the same phase distortion along the FOPA link, we included a random time shift in the pump-phase modulation sinusoidal waveform at each FOPA stage. The amplifier's noise figure was $4.5$ dB. We employed the directly-counted bit-error-rate (BER) as a system’s performance metric measured over $100\times2^{16}$ symbols. 

The kernel and the training algorithm were optimised by carrying out a grid search over the hyper-parameter space of dictionary size $M$, kernel width $\sigma$ and regularisation parameter $\lambda$ with $50\times2^{16}$ symbols generated after $10$ FOPA stages. The block length was $L = 20$. We can see from Fig.\ \ref{fig:figure2}(a) that the BER is minimised by the selection of hyper-parameters: $(M = 50,\,\sigma = 10^{0.5},\,\lambda=10^{-1})$ (marked by a blue circle). We then used these values for the performance evaluation. 

\begin{figure}[t!]
    \centering
    \includegraphics[width=7cm]{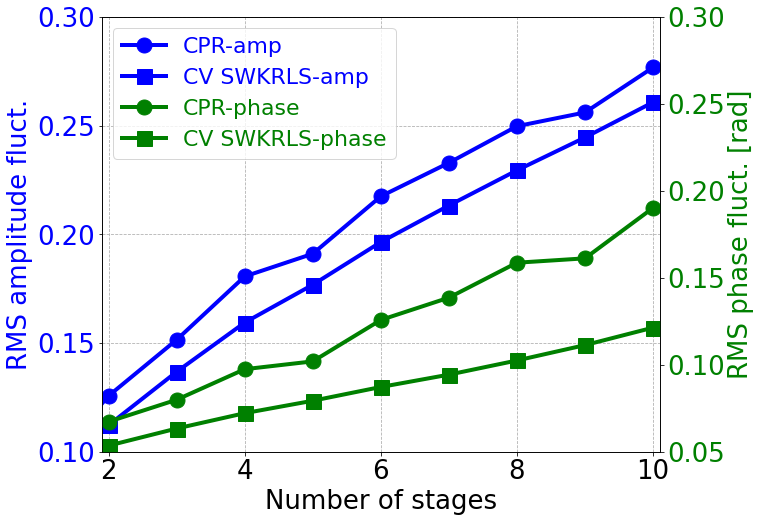}
    \caption{Amplitude and phase RMS variations over number of cascaded FOPA stages between two methods.}
    \label{fig:figure3}
\end{figure}

We compared the performance of our proposed kernel-based compensation scheme across the FOPA cascade with  conventional CPR accomplished by the one-tap least mean-squares (LMS)  algorithm\cite{fatadin_jlt2009}, as shown in Fig.\ \ref{fig:figure2}(b). Being designed to operate on the slower laser phase noise, the LMS-CPR scheme is unable to handle the high-frequency, dithering-induced phase distortions effectively, leading to phase cycle slips and increased phase detection errors. The random time shift at each stage leads to the different phase and amplitude RMS fluctuations of the received signal in different runs, and when the LMS method fails to predict it, its BER values are unstable as in Fig. \ref{fig:figure2}(b) despite of high number of symbol batches. In contrast, our kernel-based scheme significantly outperforms the LMS method, achieving a BER improvement of at least one order of magnitude with a consistent performance. The capability of the CV SWKRLS algorithm to mitigate both phase and amplitude distortions accrued along the FOPA cascade is illustrated by Fig.\ \ref{fig:figure3}. The figure shows the evolution over the transmission length of the RMS deviations of the phases and amplitudes, respectively, of the received symbols compared to the transmitted ones, after post-processing by the one-tap LMS algorithm and our method. The RMS values displayed are averaged over the alphabet size. While the suppression of phase distortion was notably effective, amplitude distortion suppression was also achieved. These results highlight the superior distortion correction capabilities of our method compared to conventional CPR.


\section{Conclusions}
We developed an kernel-based algorithm for phase and amplitude compensation in systems with multiple cascaded FOPAs. Our scheme can outperform conventional CPR schemes, which typically react slowly to temporal phase variations. The efficacy of our method has been demonstrated in 16-QAM signal transmission, achieving a BER improvement of at least an order of magnitude over the one-tap LMS phase recovery algorithm across a cascade of ten FOPAs. 


\clearpage
\section{Acknowledgements}
This work was supported by the H2020 MSCA ETN project POST-DIGITAL (EC 263 GA 860360), and the UK EPSRC grants TRANSNET (EP/R035342/1) and CREATE (EP/X019241/1).

\printbibliography

\vspace{-4mm}

\end{document}